\documentclass[12pt]{article}

\usepackage{amsmath,amssymb,amsfonts}
\usepackage{epsfig,cancel}

\usepackage{amsthm}
\usepackage[mathscr]{eucal}
\usepackage{graphicx}
\usepackage{color}
\usepackage{hypbmsec}

\numberwithin{equation}{section}

\newlength{\xtrawidth}
\newlength{\xtraheight}
\newcommand{\beq}{\begin{equation}}
\newcommand{\eeq}{\end{equation}}
\newcommand{\ba}{\begin{array}}
\newcommand{\ea}{\end{array}}
\newcommand{\bea}{\begin{eqnarray}}
\newcommand{\eea}{\end{eqnarray}}
\newcommand{\bean}{\begin{eqnarray*}}
\newcommand{\eean}{\end{eqnarray*}}
\newcommand{\eref}[1]{(\ref{#1})}

\newcommand{\comment}[1]{}

\newcommand{\cN}{{\cal N}}

\newcommand{\cA}{{\cal A}}
\newcommand{\cB}{{\cal B}}
\newcommand{\cC}{{\cal C}}

\newcommand{\cV}{{\cal V}}

\def\cjn1{{\cA, \cC^*\otimes \wedge^j \cN^*}}
\def\bjn1{{\cA, \cB^*\otimes \wedge^j \cN^*}}
\def\vjn1{{\cA, \cV^*\otimes \wedge^j \cN^*}}
\def\cjn2{{\cA, \cC\otimes \wedge^j \cN^*}}
\def\bjn2{{\cA, \cB\otimes \wedge^j \cN^*}}
\def\vjn2{{\cA, \cV\otimes \wedge^j \cN^*}}

\def\fnote#1#2{\begingroup\def\thefootnote{#1}\footnote{#2}
     \addtocounter{footnote}{-1}\endgroup}

\begin{document}

\title{{\bf Stabilizing the Complex Structure\\ in Heterotic Calabi-Yau Vacua}}

\vspace{2cm}

\author{
Lara B. Anderson${}^{1}$,
James Gray${}^{2}$,
Andre Lukas${}^{2}$,
Burt Ovrut${}^{1}$
}
\date{}
\maketitle
\begin{center} {\small ${}^1${\it Department of Physics, University of
      Pennsylvania, \\ Philadelphia, PA 19104-6395, U.S.A.}
    \\${}^2${\it Rudolf Peierls Centre for Theoretical Physics, Oxford
      University,\\
      $~~~~~$ 1 Keble Road, Oxford, OX1 3NP, U.K.}\\
 \fnote{}{andlara@physics.upenn.edu,~~j.gray1@physics.ox.ac.uk,~~lukas@physics.ox.ac.uk, ~~ovrut@elcapitan.hep.upenn.edu} 
}
\end{center}

\abstract{In this paper, we show that the presence of gauge fields in heterotic Calabi-Yau compacitifications can cause the stabilization of some, or all, of the complex structure moduli 
while maintaining a Minkowski vacuum. Certain deformations of the Calabi-Yau complex structure, with all other moduli held fixed, can lead to the gauge bundle becoming non-holomorphic and, hence, non-supersymmetric. This is manifested by a positive F-term potential which stabilizes the corresponding complex structure moduli. We use $10$- and $4$-dimensional field theory arguments as well as a derivation based purely on algebraic geometry to show that this picture is indeed correct. An explicit example is presented in which a large subset of complex structure moduli is fixed. We demonstrate that this type of theory can serve as the hidden sector in heterotic vacua
with realistic particle physics.}

\newpage

\section{Introduction}
Compactification of the $E_{8} \times E_{8}$ heterotic string \cite{Candelas:1985en,Green:1987mn} and heterotic $M$-theory \cite{Witten:1996mz,Lukas:1997fg,Lukas:1998yy} on a Calabi-Yau manifold is a compelling way to derive particle physics in four-dimensional Minkowski space~\cite{burt,Bouchard:2005ag,Anderson:2009mh}. However, moduli stabilization has been a long-standing problem in such models. The mechanisms found in flux compactifications of the IIB string cannot be directly transferred to the heterotic case since only a restricted set of fluxes, namely NS flux, is available. In particular, the stabilization of Calabi-Yau complex structure moduli in heterotic compactifications has remained problematical. In this paper, we will show that some, and possibly all, Calabi-Yau complex structure moduli can be stabilized due to the presence of gauge fields in heterotic compactifications, while maintaining a Minkowski vacuum. The basic idea behind this mechanism can be easily explained. For the internal gauge fields in a heterotic compactification to preserve supersymmetry, the associated gauge bundle has to be holomorphic and poly-stable. The condition of poly-stability and its dependence on the moduli was analysed in previous work by the authors~\cite{Anderson:2009sw,Anderson:2009nt,Anderson:2010tc}. In particular, it was found that it can lead to restrictions in the K\"ahler moduli space of the Calabi-Yau manifold and to the related  phenomenon of ``stability walls". In the present paper, we are concerned with the first condition; that is, the holomorphy of the vector bundle. Consider a holomorphic vector bundle over a Calabi-Yau manifold with a given K\"ahler class and complex structure and perform a small variation of the complex structure, keeping all other moduli fixed. The question is whether the vector bundle can re-adjust so that it stays holomorphic under the new deformed complex structure. As we will see, this is not always possible and directions in complex structure moduli space which are obstructed in this way have to be removed from the moduli space; that is, these directions have been stabilized. The superpotential vanishes on the remaining moduli space and, hence, the four-dimensional spacetime is still Minkowski. 

Let us now discuss this mechanism in more detail. It has long been known \cite{Green:1987mn} that the condition for preserving supersymmetry is that the variations of the ten-dimensional gravitino, dilatino and the gauginos vanish under supersymmetry transformations. Assuming ${\cal{N}}=1$ supersymmetry in four-dimensional Minkowski space, a constant dilaton and a vanishing harmonic part of the $H$-flux, these reduce to: 1) that the internal manifold,  $X$, be a complex, K\"ahler manifold with vanishing first Chern class, that is, a Calabi-Yau three-fold, and 2) that the gauge connection satisfy the so-called hermitian Yang-Mills equations with zero slope. 

A Calabi-Yau manifold admits an integrable complex structure $J$ whose deformations are described by $H^{1}(TX)=H^{(2,1)}(X)$. The manifold also admits a Ricci-flat K\"ahler metric, $g$. The deformations of the associated K\"ahler form are given by $H^{(1,1)}(X)$. These cohomology groups correspond to the complex structure and K\"ahler moduli respectively in the four-dimensional theory. For a fixed complex structure, which defines holomorphic and anti-holomorphic coordinates $z^{a}$ and ${\bar{z}}^{{\bar{b}}}$, and a specified K\"ahler metric $g_{a{\bar{b}}}$, the zero slope hermitian Yang-Mills equations \cite{Green:1987mn} for the gauge connection are given by 
\begin{equation}
\label{1}
g^{a{\bar{b}}} F_{a{\bar{b}}} =0 
\end{equation}
and
\begin{equation}
F_{ab} =F_{{\bar{a}}{\bar{b}}}=0 \ ,
 \label{2}
\end{equation}
where $F$ is the gauge field strength associated with a connection $A$ on a holomorphic vector bundle $V$. For a fixed complex structure, the deformations of a given gauge connection satisfying the hermitian Yang-Mills equations which preserve the holomorphy of its field strength, that is, equation (\ref{2}), are described by $H^{1}(V \otimes V^{*})$. These correspond to the vector bundle moduli in the four-dimensional theory.

The necessity for the gauge connection to satisfy the hermitian Yang-Mills equations in order to 
preserve ${\cal{N}}=1$ supersymmetry can be seen directly from effective field theory. Starting with the the ten-dimensional $E_{8} \times E_{8}$ action and using the Bianchi identity
for the antisymmetric field strength $H$, we find that
\begin{equation}
S = - \frac{1}{2\kappa_{10}^{2}} \alpha' \int_{{\cal{M}}_{10}}\sqrt{-g} \left\{\frac{1}{2} \textnormal{Tr}(g^{a{\bar{b}}} F_{a{\bar{b}}})^{2}+\textnormal{Tr}(g^{a{\bar{a}}} g^{b{\bar{b}}} F_{ab}F_{{\bar{a}}{\bar{b}}}) \right\}+\dots \ .
\label{3}
\end{equation}
\noindent Note that here, and for the remainder of the paper, we focus
only on a single $E_{8}$ gauge group for notational simplicity. All
results are easily extended to the second $E_{8}$ factor.  The terms
in (\ref{3}) form part of the ten-dimensional theory which does not
contain any four-dimensional derivatives. It therefore contributes,
upon dimensional reduction, to the potential of the four-dimensional
theory. For a gauge field configuration satisfying the hermitian
Yang-Mills equations (\ref{1}) and (\ref{2}), the terms in the
integrand vanish and no potential is generated. This corresponds, in
the four-dimensional theory, to the vanishing of both the D- and
F-terms; that is, to a supersymmetric vacuum state. Conversely,
since Eq.~\eqref{3} consists of positive definite terms,
supersymmetry in the four-dimensional theory requires that the
connection satisfies (\ref{1}) and (\ref{2}).
 
 In a series of papers \cite{Anderson:2009sw,Anderson:2009nt,Anderson:2010tc,Sharpe:1998zu}, we discussed the consequences of starting with a supersymmetric point in moduli space and then changing the vacuum by deforming both K\"ahler and vector bundle moduli {\it while keeping the complex structure fixed}. Since, by definition, this leaves the holomorphic field strength equation (\ref{2}) unchanged, such deformations can only effect $g^{a\bar{b}}F_{a{\bar{b}}}$. For vacua with $h^{(1,1)}(X)\geq 2$, the combined K\"ahler and vector bundle moduli space generically decomposes into regions where equation (\ref{1}) is satisfied and regions where it  is not; that is, ${\cal{N}}=1$ supersymmetric and non-supersymmetric regions respectively. These are separated by ``walls of stability'', where equation (\ref{1}) and, hence, supersymmetry continue to be satisfied, but additional anomalous $U(1)$ groups appear in the low energy theory. Furthermore, by analyzing the
 $\textnormal{Tr}(g^{a{\bar{b}}} F_{a{\bar{b}}})^{2}$ term in (\ref{3}), we were able to show that it is equivalent to D-term contributions to the four-dimensional potential energy, where the D-terms are 
associated with the anomalous $U(1)$ gauge factors.  
The D-terms vanish in a supersymmetric region and on a stability wall, but are non-zero and generate a positive definite potential in a non-supersymmetric region. 

Having done this, it is natural to ask what are the consequences of starting with a supersymmetric point in moduli space and then changing the vacuum with the remaining moduli; that is, deforming the complex structure moduli {\it while keeping the K\"ahler and vector bundle moduli fixed}. In this case, the definition of holomorphic and anti-holomorphic coordinates changes and equation (\ref{2}) need no longer be satisfied. Even if it is, it remains to be shown that $g^{a{\bar{b}}} F_{a{\bar{b}}}$ will vanish. In this paper, we analyze this question from three different points of view. The main conclusions are: 

\begin{itemize}
\item Given some holomorphic vector bundle, $V$, there will generically be directions in complex structure space for which equation (\ref{2}) {\it cannot} be satisfied for any deformed connection on $V$. It follows that the variation of the gauginos in those directions will no longer vanish and supersymmetry will be spontaneously broken.

\item By analyzing the ${\rm Tr}(g^{a{\bar{a}}} g^{b{\bar{b}}} F_{ab}F_{{\bar{a}}{\bar{b}}})$ term in (\ref{3}), we are able to show  that this is {\it equivalent to F-term contributions} to the four-dimensional potential energy. For complex structure deformations where equation (\ref{2}) is no longer satisfied, it follows that at least one F-term is non-vanishing; signalling a positive definite value of the potential and the breaking of supersymmetry.

\item Since the potential energy is no longer zero along these directions, the associated complex structure moduli are {\it stabilized to their initial, supersymmetric values}.
\end{itemize}

\section{10-d Field Theory}

We begin by analyzing complex structure deformations within the context of the ten-dimensional theory. Since the definition of holomorphic/anti-holomorphic coordinates changes with a deformation of the complex structure, we find it convenient to begin with an arbitrary set of real coordinates $x^{\mu}$. For a given complex structure $J$, holomorphic and anti-holomorphic objects are obtained by acting with the projection operator $P_{\mu}^{\nu}=\frac{1}{2}({ \bf{1}}_{\mu}^{\nu}+iJ_{\mu}^{\nu})$ and its conjugate
${\bar{P}}_{\mu}^{\nu}=\frac{1}{2}({ \bf{1}}_{\mu}^{\nu}-iJ_{\mu}^{\nu})$. The hermitian Yang-Mills equations can now be written as
\begin{equation}
g^{\mu\nu}P_{\mu}^{\gamma}{\bar{P}}_{\nu}^{\delta}F_{\gamma\delta}=0
\label{4}
\end{equation}
and
\begin{equation}
P_{\mu}^{\nu}P_{\rho}^{\sigma}F_{\nu\sigma}={\bar{P}}_{\mu}^{\nu}{\bar{P}}_{\rho}^{\sigma}F_{\nu\sigma}= 0 
\label{5}
\end{equation}
respectively. Fix an initial vacuum specified by a complex structure, K\"ahler class and a connection satisfying (\ref{4}) and (\ref{5}). These initial quantities will be specified by the superscript $(0)$. Now consider small deformations of the complex structure, leaving the K\"ahler class and the vector bundle moduli fixed. We will, however, allow arbitrary non-harmonic variations, $A=A^{(0)}+\delta A$, of the gauge connection to occur as we change the complex structure\footnote{Note that we refer to the bundle moduli, that is, elements $\delta A_h \in H^1(V\otimes V^*)$, as harmonic variations of the gauge field (since each equivalence class in $H^1(V\otimes V^*)$ has a harmonic representative). For {\it fixed complex structure}, these $\delta A_h$ deformations of the connection preserve $F_{ab}=0$ to linear order. By contrast, in this section we will be interested in non-harmonic changes, $\delta A$, to the connection; that is, any change in $A$ that is not an element of $H^1(V\otimes V^*)$. By definition, these $\delta A$ are not closed with respect to the gauge covariant derivative $D_{\bar{a}}$. }. Explicitly, we perturb the complex structure to $J=J^{(0)}+\delta J$  requiring that $J^{2}=-1$ and that it remains integrable. This induces the changes $P=P^{(0)}+\delta P$.

Substituting these deformations into (\ref{5}) leads to a constraint between the change in the complex structure and the deformation of the connection. Satisfying this constraint ensures that, with respect to the new complex structure, the perturbed holomorphic and anti-holomorphic gauge field strengths vanish. In analyzing this constraint, it is expedient to write the results in terms of the holomorphic and anti-holomorphic coordinates associated with the unperturbed complex structure $J^{(0)}$. Demanding that $J$ also be an integrable complex structure restricts the non-zero components of $\delta J$ to be $\delta J_{a}^{~{\bar{b}}}$ and $\delta J_{{\bar{a}}}^{~b}$ where $\delta J_{a}^{~{\bar{b}}}=-i{\bar{v}}_{I a}^{{\bar{b}}} \delta {\mathfrak{z}}^{I}$.
Here ${\bar{v}}_{I }$ are tangent bundle valued harmonic one-forms and $ \delta {\mathfrak{z}}^{I}$ are the small changes in complex structure moduli. Inserting these perturbations into (\ref{5}) gives
\begin{equation}
\delta {\mathfrak{z}}^{I} v_{I [{\bar{a}}}^{c}F_{|c|{\bar{b}}]}^{(0)} + 2D^{(0)}_{[{\bar{a}}} \delta A_{{\bar{b}}]}=0 \ ,
\label{6}
\end{equation}
where $D^{(0)}_{\bar{a}}$ is the $E_{8}$ covariant derivative with respect to $A^{(0)}_{\bar{a}}$.
The first term is the amount of the original $(1,1)$ part of the field strength that gets rotated into the 
$(0,2)$ component by changing the complex structure. The second term is the change in the initially vanishing $(0,2)$ part of the field strength due to the change in the gauge connection.

If there is {\it no} solution to (\ref{6}) for a given $\delta {\mathfrak{z}}^{I}$, then the bundle cannot adapt so as to stay holomorphic. In this case, the complex structure deformation cannot preserve supersymmetry and is not a modulus of the compactification. It follows from the positive definite ${\rm Tr}(g^{a{\bar{a}}} g^{b{\bar{b}}} F_{ab}F_{{\bar{a}}{\bar{b}}})$ term in (\ref{3}) that the associated complex structure is stabilized at its initial value. If, on  the other hand, for a given $\delta {\mathfrak{z}}^{I}$ there {\it is} a solution to (\ref{6}) for some $\delta A$, then the bundle can adapt to remain holomorphic as the complex structure of the Calabi-Yau three-fold varies. However, one must still show that equation (\ref{4}) is satisfied. This is required for  supersymmetry to be  preserved and, hence, that the resulting complex structure deformation, in combination with the change in the gauge field strength $\delta A$, be a modulus.
Perturbing (\ref{4}) to linear order, in the same way as done for (\ref{5}), gives
\begin{equation}
g^{(0)a{\bar{b}}}D^{(0)}_{[a}\delta A_{{\bar{b}}]}=0 \ .
\label{7}
\end{equation}
We now show that this equation can always be satisfied. To do this, first note that if $\delta 
{\tilde{A}} _{{\bar{a}}}$ is a solution of (\ref{6}) for given $\delta {\mathfrak {z}}^{I}$, then $\delta A_{{\bar{a}}}=\delta 
{\tilde{A}} _{{\bar{a}}}+D^{(0)}_{{\bar{a}}}\Lambda$ is also a solution for any bundle valued function $\Lambda$. It is this freedom that will always allow a solution of (\ref{7}) to exist. Substituting this expression for $\delta A_{{\bar{a}}}$ into (\ref{7}) yields the equation
\begin{equation}
g^{(0)a {\bar{b}}} \partial_{a} \partial_{\bar{b}} \Lambda+S=0, \quad S=2g^{(0)a{\bar{b}}}D^{(0)}_{[a}\delta {\tilde{A}}_{{\bar{b}}]} \ .
\label{8}
\end{equation}
An elementary result in elliptic theory tells us that this equation has a solution if and only if $S$ integrates to zero over the Calabi-Yau three-fold. This is indeed the case, see~\cite{Li:2004hx,Huybrechts}, and, hence, a solution to (\ref{8}) exists. If we further specify that $\delta A \rightarrow 0$ as $\delta{\mathfrak{z}}^{I} \rightarrow 0$, then the solution for $\Lambda$ is unique.
We conclude that if, for any $\delta {\mathfrak{z}}^{I}$, there is a deformation of the connection satisfying (\ref{6}), then there exists a unique such deformation for which equation (\ref{4}) continues to be satisfied. Hence, supersymmetry remains unbroken and these deformations correspond to a modulus of the compactification.

\section{The 4-d Field Theory}

It is useful to analyze complex structure deformations from the point of view of the four-dimensional effective field theory obtained by dimensional reduction on the Calabi-Yau three-fold $X$ with holomorphic vector bundle $V$. In particular, the field strength $H$ on $X$ contributes to the four-dimensional superpotential through the Gukov-Vafa-Witten expression \cite{Gukov:1999ya}
\begin{equation} 
 W = \int_X \Omega \wedge H \ ,
\label{4d1}
\end{equation}
where we choose
\begin{equation}
H = dB - \frac{3 \alpha'}{\sqrt{2}} \left( \omega^{3 \textnormal{YM}} - \omega^{3 {\textnormal L} }\right) \ .
\label{4d2}
\end{equation}
Here $dB$ is an {\it exact} three-form and $\omega^{3
  \textnormal{YM}}$, $\omega^{3L}$ are the gauge field and
gravitational Chern-Simons forms.  Note that $W$ is a function of the
complex structure moduli ${\mathfrak{z}}^{I}$ and fields $C_i$ descending
from the ten-dimensional gauge fields. These
parameterize the volume form $\Omega$ and $\omega^{3 \textnormal{YM}}$
respectively. In this section, we include in the $C_{i}$ fields both {\it harmonic} deformations of the gauge connection, that is, the vector bundle moduli, as well as {\it non-harmonic} deformations. We emphasize that the $dB$ term in \eqref{4d2} is {\it
  global} and, hence, our discussion is entirely within the context of
a complex, K\"ahler, Calabi-Yau three-fold, albeit one with
non-vanishing Ricci tensor at order $\alpha'$. Were one to allow a
non-vanishing harmonic flux $H_{0}$, that is, one for which $[H_{0}]
\neq 0$, then this ``flux vacuum'' would no longer be Calabi-Yau.

It is well-known that the holomorphic gauge field strength $F_{ab}$
dimensionally reduces to the $\partial W / \partial C_i$ terms in
the four-dimensional theory. Recall that we are restricting our
discussion to Minkowski space. 
Hence, the vacuum value of $W$
vanishes. Therefore, the relevant $F$-terms become
\begin{equation}
F_{C_i}=\frac{\partial W}{\partial C_i} =-\frac{3 \alpha'}{\sqrt{2}} \int_X \Omega \wedge \frac{\partial
 \omega^{3 \textnormal{YM}}}{\partial {C_i}} \ .
\label{4d3}
\end{equation}
We have used the fact that only $\omega^{3 \textnormal{YM}}$ depends on the fields $C_i$. For any initial complex structure ${\mathfrak{z}}^{(0)I}$ for which the connection $A^{(0)}$ is holomorphic and supersymmetric, all 
$F_{C_i}=0$. Now vary the complex structure, and, hence, $\Omega$, by $\delta {\mathfrak{z}}^{I}$ and perturb
\begin{equation}
A_{\mu} = A^{(0)}_{\mu} + \delta A_{\mu}+\bar{\omega}^{i}_{\mu}
\delta C_{i} + \omega^{i}_{\mu}\delta{ \bar{C}}_{i} \ .
\label{4d4}
\end{equation}
We allow the $\bar{\omega}$ to be both harmonic  and non-harmonic forms with respect to the
background connection $A^{(0)}$ and $\delta C_{i}$ are variations of
the $C_i$.  Evaluating \eqref{4d3} to linear order in
$\delta {\mathfrak{z}}^{I}$ and $\delta A$, we find
\begin{equation}
F_{C_i} = \int_X \epsilon^{\bar{a}\bar{c}\bar{b}}
\epsilon^{abc} \Omega^{(0)}_{abc} 2 \bar{\omega}^{x i}_{\bar{c}} \,
\textnormal{tr}(T_{x}T_{y})\left( \delta {\mathfrak{z}}^{I} v_{I [{\bar{a}}}^{c}F_{|c|{\bar{b}}]}^{(0)y} + 2D^{(0)}_{[{\bar{a}}} \delta A^{y}_{{\bar{b}}]} \right) 
\label{4d5}
\end{equation}
where $T_{x}$ are the generators of $E_{8}$.

Note that for complex structure deformations $\delta
{\mathfrak{z}}^{I}$ for which {\it there exists} $\delta A$ satisfying
$\delta {\mathfrak{z}}^{I} v_{I [{\bar{a}}}^{c}F_{|c|{\bar{b}}]}^{(0)}
+ 2D^{(0)}_{[{\bar{a}}} \delta A_{{\bar{b}}]}=0$, all $F_{C_i}$ terms {\it vanish}. It follows that these
deformations are not obstructed by the potential energy and, hence,
are complex structure moduli.  On the other hand, for deformations
$\delta {\mathfrak{z}}^{I}$ for which there is {\it no} $\delta A$
which sets the integrand to zero, at least one $F_{C_i}$
term is {\it non-vanishing}. The corresponding complex structure
deformations are then obstructed by a positive potential and, hence,
these fields are massive and fixed at their initial value.  To be
accurate, this last statement is only suggestive. The correct
statement is that, in fact, both the complex structure deformations
obstructed in ten-dimensions, and some of the fields $C_i$, specifically, the non-harmonic modes, are generically not
zero-modes of the respective Dirac operators. Hence, they would not
normally be regarded as fields in the four-dimensional effective
theory. Be that as it may, one can see that modes obstructed in
ten-dimensions, if viewed from the dimensionally reduced
four-dimensional theory, will appear in non-vanishing $F$-term
contributions to the potential energy. This gives them a mass
determined in terms of $F^{(0)}_{a\bar{b}}$ and, hence, stabilizes their
values. Generically, this mass will be of the same order as other
heavy states descending from the gauge fields and, hence, the analysis
is only suggestive. There are examples, however, when this mass,
although non-zero, will be much smaller than other mass scales \cite{big_paper}.
In these cases, the four-dimensional discussion of fixing
some complex structure moduli is valid.

\section{Algebraic Geometry -- The Atiyah Class}\label{atiyah_sec}

We now present a third approach to analyzing complex structure deformations purely within the context of algebraic geometry. Begin by defining an initial Calabi-Yau three-fold $X$ and a holomorphic vector bundle $V$ over $X$ by specifying the complex structure, K\"ahler form and connection  respectively. Of interest is the space of simultaneous {\it holomorphic} deformations of $X$ and $V$. The associated tangent space  was introduced by Atiyah \cite{Atiyah}.  It is given by $H^{1}({\cal{Q}})$, where $\cal{Q}$ is defined by the short exact extension sequence
\begin{equation}
0 \to V\otimes V^* \to {\cal Q} \stackrel{\pi}{\to} TX \to 0 
\label{9}
\end{equation}
and, additionally, the extension class is chosen to be
\begin{equation}
\alpha=[F^{(0)1,1}] \in H^1(V\otimes V^* \otimes TX^*)\; ,
\label{10}
\end{equation}
that is, by the $(1,1)$ component of the initial field strength. The class $\alpha$ is referred to as the ``Atiyah class''. Associated with (\ref{9}) is a long exact sequence in cohomology. Since $TX$ is a stable bundle, $H^0(TX)=H^3(TX)=0$, and, hence, the long exact sequence takes the form
\begin{equation}
0 \to H^1(V\otimes V^*) \to H^1({\cal Q}) \stackrel{d\pi}{\to} H^1(TX) \stackrel{\alpha}{\to} H^2(V\otimes V^*) \to \ldots  \ .
\label{11}
\end{equation}

Note that since $H^1(V\times V^*)$ injects into $H^1({\cal Q})$, the familiar vector bundle moduli are a subspace of $H^1({\cal Q})$. What about the complex structure moduli? If the map $d\pi$ from $H^1({\cal Q})$ to $H^1(TX)$ is surjective, the sequence (\ref{11}) splits, the relevant part being 
\begin{equation}
0 \to H^1(V\otimes V^*) \to H^1({\cal Q}) \stackrel{d\pi}{\to} H^1(TX) \stackrel{\alpha}{\to} 0 \ .
\label{12}
\end{equation}
Then $H^1({\cal Q})$ is just the direct sum $H^1(V\otimes V^*) \oplus H^1(TX)$. In this case, it follows that for each complex structure deformation, the vector bundle remains holomorphic. However, the map $d\pi$ need {\it not} be surjective. Then all one can say is that
\begin{equation}
H^1({\cal Q})=H^1(V\otimes V^*) \oplus {\rm Im}(d\pi) \ ,
\label{13}
\end{equation}
where ${\rm Im}(d\pi)$ is some proper subset of the complex structure moduli space $H^1(TX)$. In this case, there exists some complex structure deformations for which the vector bundle cannot remain holomorphic.

It is difficult to formulate the map $d\pi$ explicitly since we have defined ${\cal Q}$ itself indirectly. However, from the exactness of sequence (\ref{11}) it follows that ${\rm Im}(d\pi)={\rm Ker}(\alpha)$. Thus, one can determine the properties of $d\pi$ by considering the map $\alpha$ defined in (\ref{10}). We conclude that the complex structure fluctuations around the initial vacuum are restricted to those elements $\nu \in H^1(TX)$ for which
\begin{equation}
\alpha(\nu)=0 \in H^2(V\otimes V^*) \ .
\label{14}
\end{equation}
If $\alpha(\nu) \neq 0$, then $\nu$ is not an allowed modulus of the theory. That is, the non-zero image of $\alpha$ in $H^2(V\otimes V^*)$ corresponds to non-vanishing holomorphic field strengths $F^{0,2}$. It follows that deformations in those directions are non-supersymmetric and, from (\ref{3}), have positive definite potential. Hence, the non-zero elements of ${\rm Im}(\alpha)$ are in one to one correspondence with the complex structure moduli that are stabilized at their initial values. Furthermore, note that ${\rm Im}(\alpha)$ is bounded by the dimension of $H^2(V\otimes V^*)$. It follows that requiring a bundle $V$ to be holomorphic, can stabilize a maximum of $h^2(V\otimes V^*)$ complex structure moduli.

The content of equation (\ref{14}) can be made more explicit by writing a general element of $H^{1}(TX)$ as $\nu=\delta {\mathfrak{z}}^{I} v_{I {\bar{a}}}^{c}$ and recalling that $\alpha=[F^{(0)1,1}]$.
Then the condition $\alpha(\nu)=0 \in H^2(V\otimes V^*)$ can be written as
\begin{equation}
\delta{\mathfrak z}^{I}v^{c}_{I[{\bar a}}F^{(0)}_{|c|{\bar b}]}=D^{(0)}_{[{\bar a}}\Lambda_{{\bar b}]} \ .
\label{15}
\end{equation}
We have used the fact that a trivial image of $\alpha$ in $H^2(V\otimes V^*)$ is, by definition, an exact bundle-valued two-form. Note that by taking $\Lambda=-2\delta A$, we {\it recover the expression (\ref{6}) derived from ten- and four-dimensional field theories}. Thus, the Atiyah class locally measures which deformations of the complex structure and connection can keep $V$ holomorphic. That is, ${\rm Ker} (\alpha)$ determines the directions in complex structure moduli space $H^{1}(TX)$ for which it is possible to satisfy $F_{ab}=F_{{\bar a}{\bar b}}=0$.

Recall that it is still necessary to show that for these deformations the field strength continues to satisfy 
$g^{a{\bar{b}}} F_{a{\bar{b}}} =0 $.
From the point of view of algebraic geometry, there are two powerful theorems that guarantee that this will always be the case. 

\begin{itemize}

\item It is known, see~\cite{Huybrechts}, that if a holomorphic
  vector bundle is {\it poly-stable} with respect to an initial complex
  structure and connection, then it remains poly-stable under any
  complex structure deformations that leave the bundle
  holomorphic. The property of poly-stability is said to be {\it open}
  in complex structure moduli space.

\item It was proven in the classic work of Donaldson \cite{duy2} and Uhlenbeck and Yau \cite{duy1} that if for fixed K\"ahler and bundle moduli a holomorphic bundle is poly-stable, then there exists a unique holomorphic connection for which $g^{a{\bar{b}}} F_{a{\bar{b}}} =0$ is satisfied.

\end{itemize}

\noindent We conclude that these results in algebraic geometry are completely equivalent to those derived above using ten- and four-dimensional field theory.

\section{An Explicit Example}

The above arguments each indicate that some, or all, of the complex
structure moduli will be stabilized if the vector bundle cannot remain
holomorphic under the associated deformations.  In the ten- and
four-dimensional field theories, this is expressed through the
equation $\delta {\mathfrak{z}}^{I} v_{I
  [{\bar{a}}}^{c}F_{|c|{\bar{b}}]}^{(0)} + 2D^{(0)}_{[{\bar{a}}}
\delta A_{{\bar{b}}]}=0$. {\it If for some $\delta {\mathfrak{z}}^{I}$
  this equation has no solution}, then the complex structure is
stabilized in those directions. Equivalently, in the algebraic
geometry approach, {\it if the Atiyah class is such that ${\rm Im} (\alpha)
  \neq 0$}, then the complex structure moduli for which $\alpha
(\delta\mathfrak{z}^{I}v_I) \neq 0$ will be stabilized. However, it is
essential to show that this possibility can indeed occur, and to
compute the number of stabilized moduli. Unfortunately, it is often
difficult to directly solve the deformation equation \eqref{6} or,
equivalently, to find ${\rm Im}(\alpha)$ since we do not know the explicit
form~\cite{Anderson:2010ke} of $F_{a{\bar{b}}}^{(0)}$. In this section, we proceed by
exploiting a class of examples with a particular algebraic
property. This property makes the holomorphic structure of the bundle
self-evident. This will allow us to solve \eref{6} without having to
specifying the gauge field strength. The following is an explicit example of such a vacuum.

Consider the complete intersection Calabi-Yau three-fold defined by
\begin{eqnarray}
\label{eg1cy3}
X=\left[ \ba{c |c }
\mathbb{P}^1 & 2 \\
\mathbb{P}^1 & 2 \\
\mathbb{P}^2 & 3
\ea \right]^{3,75} \ .
\end{eqnarray}
As indicated, the number of K\"ahler moduli is $h^{1,1}(X)=3$ and the number of complex structure moduli is $h^{1}(TX)=h^{1,2}(X)=75$. We denote these as $t^{r}$ and ${\mathfrak{z}}^{I}$ respectively. Note that the K\"ahler cone for this example is the positive octant of ${\mathbb{R}}^{3}$. Over this manifold, we want to define a {\it holomorphic,  indecomposable} $SU(2)$ vector bundle, $V$, by extension of a line bundle ${\cal{L}}={\cal{O}}_{X}(-2,-1,2)$ and its dual. That is,
\begin{equation}
0 \longrightarrow {\cal L} \longrightarrow V \longrightarrow {\cal L}^*\longrightarrow 0 \ .
\label{17}
\end{equation}
To define an indecomposable $SU(2)$ bundle, the extension class of $V$ must be a non-zero element of ${\rm Ext}^{1}({\cal{L}}^{*},{\cal{L}})=H^{1}(X,{\cal{L}}^{2})$. 
However, direct computation at a {\it generic} point ${\mathfrak{z}}^{I}$ in $H^{1}(TX)$ gives
\begin{equation}
H^{1}(X,{\cal{L}}^{2})=H^{1}({\cal{O}}_{X}(-4,-2,4))= 0. 
\label{18}
\end{equation}
Hence, for a {\it generic} complex structure a holomorphic, indecomposable 
$SU(2)$  bundle $V$ cannot be defined. Instead, the only holomorphic bundle available is the ``split extension"; that is, the direct sum
\begin{equation}
{\cal{L}} \oplus {\cal{L}}^{*}={\cal{O}}_{X}(-2,-1,2) \oplus  {\cal{O}}_{X}(2,1,-2) 
\label{19}
\end{equation}
with the reducible structure group $S[U(1) \times U(1)]$. 
We note that such a bundle can only be poly-stable and, hence, supersymmetric on the special locus in K\"ahler moduli space where the slopes $\mu({\cal{L}})=\mu({\cal{L}}^{*})=0$. This occurs when $6t^{1}t^{2}=(2t^{2}+3t^{3}+2t^{1})t^{3}$. Away from this locus, the bundle in (\ref{19}) is unstable and breaks supersymmetry. 

Despite the fact that (\ref{18}) holds generically, it is well-known that cohomology can ``jump'' on certain higher co-dimensional loci in complex structure moduli space \cite{Anderson:2009mh,Anderson:2009ge,moduli,Braun:2005xp}. Using the long exact sequence associated with the Koszul complex for ${\cal{L}}={\cal{O}}_{X}(-2,-1,2)$ and the Bott-Borel-Weil polynomial representations for the ambient projective space cohomology groups \cite{Anderson:2007nc,Anderson:2008uw,Anderson:2008ex},  one can compute the extension space at any fixed value of complex structure. Direct calculation based on~\cite{Gray,cicy} demonstrates that on a {\it $58$-dimensional sub-locus of $H^{1}(TX)$} the extension space becomes non-vanishing with dimension
\begin{equation}
h^{1}(X,{\cal{L}}^{2})=h^{1}({\cal{O}}_{X}(-4,-2,4))= 18 \ .
\label{21}
\end{equation}
That is, at any point on this sub-locus there are {\it non-zero} extension classes and, hence, a {\it holomorphic, indecomposable bundle $V$ with irreducible $SU(2)$ structure group} can be defined. 
As discussed in~\cite{Anderson:2009sw,Anderson:2009nt}, for an arbitrary bundle the K\"ahler cone can break into regions where the bundle is stable, regions where it is  unstable and ``walls of stability'', where the bundle can be poly-stable, separating them. For any indecomposable $SU(2)$ bundle defined by \eqref{17}, the region of stability is given by $6t^{1}t^{2} < (2t^{2}+3t^{3}+2t^{1})t^{3}$.

Using these results, we can present our explicit example. Begin by choosing the initial complex structure moduli at a point ${\mathfrak{z}}^{(0)I}$ on the $58$-dimensional sub-locus where \eqref{21} holds. Next, take $V$ to be defined by a non-zero extension class of \eqref{21} {\it far} from the zero class; that is, with large vector bundle moduli ${C}^{(0)}_{i}$. Furthermore, choose the initial K\"ahler moduli $t^{(0)r}$ to be deep in the stable region of $V$, far from the ``stability wall" at $6t^{1}t^{2} = (2t^{2}+3t^{3}+2t^{1})t^{3}$. Note that these two conditions guarantee that the supersymmetric connection $A^{(0)}_{\bar{b}}$ on the indecomposable $SU(2)$ bundle is {\it not infinitesimally close} to the reducible $S[U(1) \times U(1)]$ valued connection of ${\cal{L}} \oplus {\cal{L}}^{*}$ in \eref{19}. Rather, its ``off-diagonal'' $SU(2)$-valued components are large and cannot be set to zero by a small change $\delta A_{\bar{b}}$. Henceforth, we hold  the extension class of $V$, the vector bundle moduli and the K\"ahler moduli fixed.

Now consider a small deformation $\delta{\mathfrak{z}}^{I}$ of the complex structure so that ${\mathfrak{z}}^{(0)I}+\delta{\mathfrak{z}}^{I}$ is a generic point {\it not contained} in the $58$-dimensional sub-locus. Note that there are $75-58=17$ independent deformations of this type.
Since at a generic point $H^1(X,{\cal L}^2)=0$, the only bundle holomorphic with respect to the new complex structure 
must be a direct sum of line bundles of the form \eref{19} with a reducible $S[U(1) \times U(1)]$ structure group. However, we have ensured that no indecomposable connection on our initial $SU(2)$ bundle, \eref{17}, can ever be transformed into a reducible one by a small deformation $\delta A_{\bar{b}}$. Therefore, for any $\delta{\mathfrak{z}}^{I}$ in one of these $17$ directions, {\it no small deformations $ \delta A_{\bar{b}}$ are possible that preserve the holomorphy of the gauge field strength}; that is,  there is no  $\delta A_{\bar{b}}$ solving equation \eqref{6}. We conclude that the complex structure moduli {\it are stabilized} in each of these $17$ directions.
Another way to see this is as follows. Recall that we have held the K\"ahler moduli fixed under the above complex structure deformation. Since the K\"ahler moduli have been chosen to be far from the stability wall, it follows that the reducible sum of line bundles in \eref{19} is unstable. However,
the ``open'' property of poly-stability \cite{Huybrechts} then guarantees that no deformation $\delta A_{{\bar{b}}}$ of the initial connection satisfying $\delta {\mathfrak{z}}^{I} v_{I [{\bar{a}}}^{c}F_{|c|{\bar{b}}]}^{(0)} + 2D^{(0)}_{[{\bar{a}}} \delta A_{{\bar{b}}]}=0$ can exist. If it did, the new holomorphic bundle would have to be stable. That is, both the D- and F-term contributions to potential \eqref{3} become positive definite. Hence, the complex structure moduli in those $17$ directions {\it are stabilized}.

What happens for deformations of the complex structure such that ${\mathfrak{z}}^{(0)I}+\delta{\mathfrak{z}}^{I}$ is {\it contained} the $58$-dimensional sub-locus? With respect to this new complex structure, the extension space of \eqref{17} remains non-vanishing with dimension $18$. Any non-zero element is a holomorphic, indecomposable bundle with irreducible $SU(2)$ structure group. In particular, this space contains a non-vanishing extension class with the same vector bundle moduli as the original bundle at ${\mathfrak{z}}^{(0)I}$. Since we have kept the K\"ahler moduli fixed, this extension class continues to be slope stable and, hence, admits a unique supersymmetric connection. By construction, this connection will be $A^{(0)}_{\bar{b}}+\delta A_{\bar{b}}$, where $\delta A_{\bar{b}}$ is a small non-harmonic form. Therefore, for any $\delta{\mathfrak{z}}^{I}$ in these $58$ directions, we would 
{\it expect} that a small deformation $ \delta A_{\bar{b}}$ exists which preserves the holomorphy of the gauge field strength.
To {\it prove} this, consider the image of the Atiyah class \eref{10} defined with respect to the initial bundle $V$. Recall from Section 4 that $\alpha:H^1(TX) \to H^2(V\otimes V^*)$ and that the number of stabilized complex deformations is given by the dimension of ${\rm Im}(\alpha)$.
Since ${\mathfrak{z}}^{(0)I}$  is on the $58$-dimensional locus for which $h^1(X,{\cal L}^2)=18$, it is straightforward to show \cite{Anderson:2009nt,Anderson:2010tc} that for $V$ in \eref{17}
\beq
h^1(V\otimes V^*)=h^1(X,{\cal L}^2)-1 \ .
\label{22}
\eeq
It follows from Serre duality that $h^1(V\otimes V^*)=h^2(V\otimes V^*)=17$ and, hence,
\beq
{\rm Im}(\alpha) \leq h^2(V\otimes V^*)=17 \ .
\label{23}
\eeq
Since we have already shown that the ``jumping" of the extension cohomology $H^1(X,{\cal L}^2)$ stabilized $17$ moduli, we see that, as expected, all complex structure deformations within the 
$58$-dimensional sub-locus are unobstructed.
That is, for these $\delta{\mathfrak{z}}^{I}$ directions there is a $\delta A_{\bar{b}}$ solving equation \eqref{6}. 
We conclude that in this example we have stabilized {\it exactly} $17$ complex structure moduli.

\section{Fixing Complex Structure in Realistic Vacua}

In this section, we analyze the possibility of stabilizing complex structure moduli by appropriately choosing the holomorphic vector bundle in the {\it hidden sector}. This approach is very appealing, since it would fix the complex structure {\it without putting serious constraints on the choice of vector bundle in the observable sector}. That is, the search for a visible sector bundle leading to realistic phenomenology would be essentially unimpeded.

Let us begin by using the explicit example we just presented for both the Calabi-Yau three-fold and the vector bundle in the hidden sector. As above, pick and fix a point in complex structure, vector bundle and K\"ahler moduli space where the holomorphic bundle is indecomposable with structure group $SU(2)$ and is supersymmetric.  This stabilizes $17$ out of the $75$ complex structure moduli of the Calabi-Yau three-fold. We note that the four-dimensional gauge group of the hidden sector is $E_{7}$. Furthermore, computing the zero mode spectrum we find only the $E_{7}$ gauge multiplet and vector bundle moduli. There are {\it no charged matter fields}. It follows that the hidden sector gauge theory will become strongly coupled at a high scale, allowing for spontaneous supersymmetry breaking via gaugino condensation.

 Let us now analyze the consequences of trying to embed this as the hidden sector of a heterotic vacuum. It is important to recognize that although this is a ``hidden'' sector, it does directly effect the observable sector through the anomaly cancellation condition. This is given by
\begin{equation}
c_{2}(X)-c_{2}(V_{\rm visible})-c_{2}(V_{\rm hidden})=[W]_{M_{5}} \ ,
\label{22}
\end{equation}
where $c_{2}$ denotes the second Chern class of the Calabi-Yau three-fold and the visible/hidden sector bundles respectively, while $[W]_{M_{5}}$ is the effective class of the holomorphic curve of wrapped five-branes. This means we can phrase the anomaly cancellation condition as
\begin{equation}
(c_{2}(X)-c_{2}(V_{\rm visible})-c_{2}(V_{\rm hidden}))_r \geq 0 
\label{23}
\end{equation}
where $r=1,\ldots ,h^{(1,1)}(X)$ labels the harmonic $(2,2)$ forms of $X$. For our explicit example, one can compute both $c_{2}(X)$ and $c_{2}(V_{\rm hidden})$.
For the Calabi-Yau manifold defined in \eqref{eg1cy3}, we find $c_{2}(X)_{r}= (24,24,36)$. Choosing $V_{\rm hidden}$ to be an $SU(2)$ bundle defined by \eqref{17}, and recalling that $c_{1}({\cal{L}})=(-2,-1,2)$, it follows that $c_{2}(V_{\rm hidden})_{r}= (4,16,12)$. Inserting these results into \eqref{23}, we conclude that the visible sector vector bundle is constrained to satisfy
\begin{equation}
c_{2}(V_{\rm visible})_{r} \leq (20,8,24) \ .
\label{27}
\end{equation}
The large integers defining the upper bound on $c_{2}(V_{\rm visible})$ mean that there is considerable freedom in choosing the observable sector bundle. In particular, one can exploit this freedom to search for a bundle $V_{\rm visible}$ with a realistic low energy gauge group and spectrum. Unfortunately, for this particular Calabi-Yau three-fold, no heterotic standard model bundles are known. However, this example clearly demonstrates the concept of fixing complex structure in the hidden sector of a heterotic vacuum. 

Can one generalize this idea so as to combine it with a realistic observable sector? For a given Calabi-Yau three-fold, there may be many different bundles whose holomorphic structure will stabilize many, or all, of the complex structure moduli. An analysis of an individual three-fold may reveal entire classes of bundles well-suited to this role. However, we point out that the $SU(2)$ extension bundles discussed above are a good, and universally available, choice for such hidden sector bundles. In particular:

\begin{itemize}

\item Non-trivial $SU(2)$ extension bundles of the form $0 \longrightarrow {\cal{L}} \rightarrow V \rightarrow {\cal{L}}^{*} \rightarrow 0$ can be defined for any Calabi-Yau three-fold with 
$h^{(1,1)}(X) \geq 2$. For appropriate choices of line bundle $ {\cal{L}}$, the defining ${\rm Ext}^{1}$ group can be chosen to ``jump'' as discussed in the last section. That is, it is generically possible to find an $SU(2)$ extension bundle whose holomorphic structure explicitly depends on the complex structure moduli.

\item As long as $H^{0}(X,{\cal{L}})=H^{3}(X,{\cal{L}})=0$, there will exist a region in K\"ahler moduli space for which $V$ is slope-stable \cite{big_paper} and, hence, supersymmetric.

\item To ensure that $H^{0}(X,{\cal{L}})=H^{3}(X,{\cal{L}})=0$, both ${\cal{L}}$ and its dual are chosen to be {\it not} ample. Hence, its first Chern class will generically be a vector with mixed positive and negative entries. For instance, in our explicit example $c_{1}({\cal{L}})=(-2,-1,2)$. As a result $c_{2}(V_{\rm hidden})$ is generally small. It follows that there is a wide range of visible sector bundles $V_{\rm visible}$ satisfying the bound \eqref{23} required by anomaly cancellation.

\item The four-dimensional gauge group of the hidden sector will be $E_{7}$. Because the associated line bundles are mixed, it will often be possible to choose examples with few, or no, charged matter multiplets in the spectrum. Hence, these hidden sectors can be asymptotically free and exhibit spontaneous supersymmetry breaking through gaugino condensation.

\end{itemize}

We conclude that stabilizing the complex structure of realistic heterotic vacua via an appropriate vector bundle in the hidden sector is a very promising scenario. It is likely that hidden bundles can be constructed which stabilize all Calabi-Yau complex structure moduli. Combining this with the K\"ahler moduli and dilaton dependence of bundle stability, as well as the associated phenomenon of ``stability walls", it would then be possible to fix all geometric moduli except one, while maintaining a supersymmetric Minkowski vacuum. This final geometric modulus, and the gauge bundle moduli, can presumably be stabilized by non-perturbative effects \cite{non-pert}. Such a complete stabilization scenario for heterotic Calabi-Yau compactifications is currently under investigation~\cite{scenario}.

\end{document}